\documentclass[journal]{IEEEtran}

\usepackage{amsmath,amssymb}
\usepackage{multicol}
\usepackage{graphicx}
\usepackage{amsmath}
\usepackage{caption}
\usepackage{algorithm}
\usepackage{algpseudocode} 
\usepackage{placeins}
\usepackage{comment}
\usepackage{float}
\usepackage{afterpage}
\usepackage{stfloats}

\usepackage{color}

\usepackage{xcolor}

\title{The Coherent Polynomials Closed-Form Model for Evaluating Nonlinear Interference in Any Island}

\author{ Yifeng~Gao, Yanchao~Jiang, Pierluigi~Poggiolini}

\begin{document}

\maketitle
\begin{abstract}
We improve the accuracy of the GN Polynomial Closed-Form Model (PCFM) by incorporating the spectral NLI PSD and the coherent accumulation along the link. The proposed model is capable of accurately evaluating the NLI over any rectangular integration domain without relying on the machine-learning correction factors, even in low-dispersion and low-baud-rate subcarrier systems. 
\end{abstract}
    
  
\section{Introduction}
\label{sec:Introduction}
Accurate modeling of nonlinear interference (NLI) is essential for the design and optimization of modern optical fiber systems, especially in emerging scenarios such as ultra-wideband, low-dispersion, and multi-subcarrier systems. While the Gaussian Noise (GN) and Enhanced GN (EGN) models provide reliable references, their numerical evaluation is computationally demanding and often unsuitable for real-time applications. To overcome this limitation, a variety of closed-form models (CFMs) have been proposed, enabling fast and scalable NLI estimation. Among them, the Polynomial Closed-Form Model (call it PCFM1) \cite{2025_JLT_Poggiolini} has recently emerged as a flexible framework, where the effective spatial power profile (SPP) is represented through a polynomial expansion, allowing the core GN-model integrals to be evaluated analytically. 

However, most existing CFMs, including the PCFM1, rely on a set of simplifying assumptions. In particular, they typically: i) evaluate the NLI power spectral density (PSD) only at the center frequency of the channel under test (CUT) and assume it to be spectrally flat; ii) approximate the GN-model integration islands with rectangles or stripes; iii) neglect multi-channel interference (MCI); and iv) assume incoherent or crudely model the coherent accumulation of NLI across spans. While these approximations are acceptable in high-dispersion and high-symbol-rate systems, the models become increasingly inaccurate in scenarios of practical interest, such as low-dispersion transmission and digital subcarrier systems, where both MCI contributions and coherence effects become significant.

In this work, we propose an upgraded version of PCFM1, which we refer to as PCFM2, that removes the above limitations within a unified analytical framework. Starting from the whole-link GN-model formulation, we introduce an accumulated-dispersion representation that explicitly captures the phase evolution along the link. This leads to a natural decomposition of the NLI contributions into intra-span and inter-span terms, both described by a common analytical structure. By representing the effective SPP with polynomials, the GN-model integrals are reduced to closed-form over arbitrary rectangular domains. The proposed model provides an accurate and computationally efficient tool for analyzing NLI accumulation in practical systems, without relying on the machine-learning correction factor(MLCF).

\section{NLI PSD}
\label{sec:NLI_PSD}
The GN model reference formula in \cite{2012_JLT_Poggiolini} gives the NLI PSD at an arbitrary frequency $f$ in CUT,
denoted by $G_{\mathrm{NLI}}(f)$,  of all heterogeneous spans along the entire link, rather than of an individual span:
\begin{equation}
    \begin{aligned}
      G_{\text{NLI}}\left( f \right) & =\frac{16}{27}{{\Gamma }}\left( f \right) \int_{-\infty }^{\infty }{\int_{-\infty }^{\infty }}\\ 
      & G_{\text{WDM}}\left( {{f}_{1}} \right) G_{\text{WDM}}\left( {{f}_{2}} \right) G_{\text{WDM}}\left( {{f}_{1}}+{{f}_{2}}-f \right) \\
      & \cdot {{{\gamma }}^{2}\left( {{f}_{1}},{{f}_{2}},f, z \right)} \cdot {{\left| \rho  \left( {{f}_{1}},{{f}_{2}},f,z \right) \right|}^{2}}d{{f}_{1}}d{{f}_{2}} \\ 
    \\
    { \rho  \left( {{f}_{1}},{{f}_{2}},f, z \right) } &= {{ {{e}^{-\int_{0}^{L_{\text{tot}}}{{{\alpha_T }}\left( f,z \right)dz}}}\int_{0}^{L_{\text{tot}}}{{{e}^{\int_{0}^{z}{\Delta {{\kappa }}\left( {{f}_{1}},{{f}_{2}},f,z' \right) d{z}'}}}}dz }}\\ 
    \end{aligned}
    \label{eq:GNLI_ns}
\end{equation}

We do not impose the identical-span assumption, but instead retain the full longitudinal dependence inside the link function $\rho(f_1,f_2,f,z)$ and the nonlinear coefficients $ {{\gamma }}\left( {{f}_{1}},{{f}_{2}},f, z \right)$ in Eq.~(\ref{eq:GNLI_ns}).  $\Gamma(f)$ denotes the accumulated gain/loss at the frequency $f$ along the entire link. The double integral over the entire $(f_1, f_2)$ plane can be naturally decomposed into multiple integration `islands' by the support of the three WDM spectral terms $G_{\text{WDM}}(f_1)$, $G_{\text{WDM}}(f_2)$, and $G_{\text{WDM}}(f_1+f_2-f)$. 

\begin{figure}[t]
    \centering
    \includegraphics[width=0.45\textwidth]{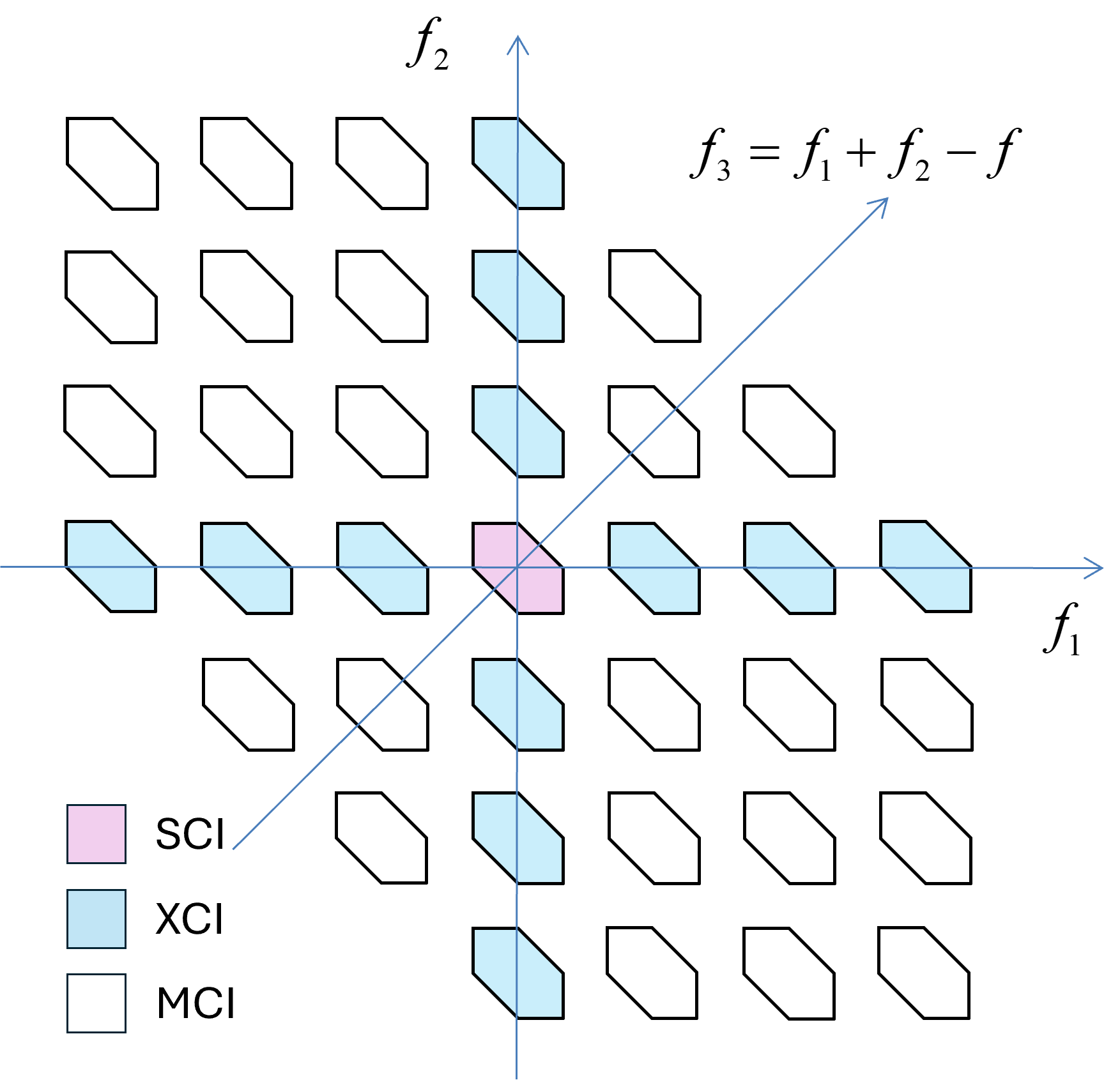}
    \caption{GN-model integration islands for the calculation of the NLI PSD at the center frequency of the center channel of a WDM system with 7 identical and equally spaced channels.}
    \label{fig:islands}
\end{figure}

\begin{figure}
    \centering
    \includegraphics[width=0.45\textwidth]{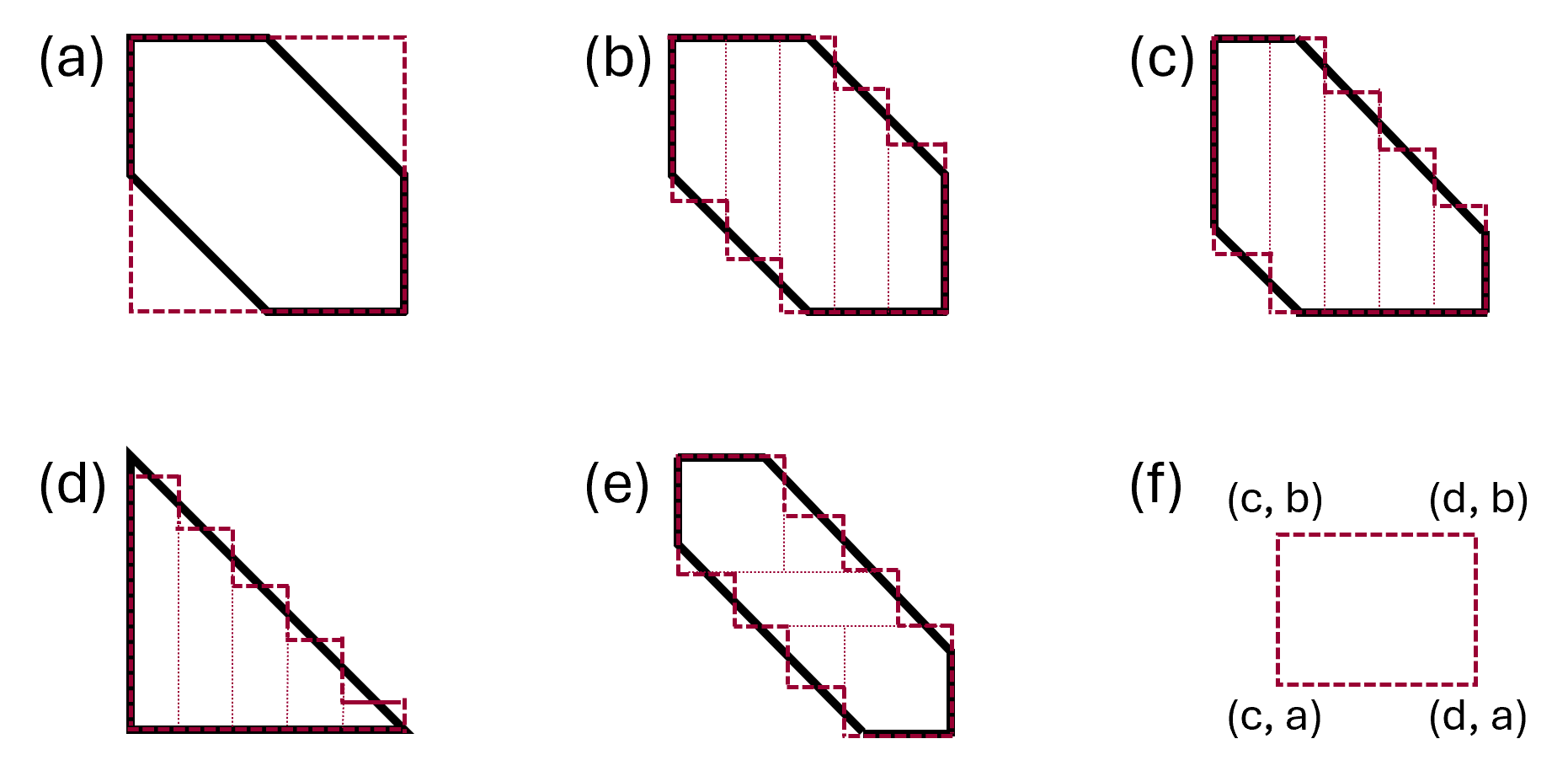}
    \caption{Black lines: Some typical GN-model integration islands. (a) typical lozenge, with the square approximation available in PCFM1 for SCI; (b) typical lozenge covered with rectangles; (c), (d), different island shapes emerging when the NLI spectrum is calculated at a frequency different from the center of a channel, and/or (e) when the WDM comb channels are different and not equally spaced; (f) an arbitrary rectangle with boundaries $a \le f_2 \le b, c \le f_1 \le d$.}
    \label{fig:arbitrary_rectangle}
\end{figure}

In detail, Fig.\ref{fig:islands} shows an example of the lozenge-shaped GN-model integration islands, for a 7-channel WDM system, when calculating the NLI spectrum at the center frequency of the center channel. SCI, XCI and MCI relate to  self-, cross- and multi-channel interference. The islands vary with the frequency $f$ at which the NLI PSD is evaluated. PCFM1 provided a closed-form solution for a square approximation of the SCI island, as shown in Fig.\ref{fig:arbitrary_rectangle}(a). The domain overestimation is mitigated by the MLCF, which works well in high dispersion only. 
When calculating the NLI PSD at frequencies other than the center of a channel, shapes like Fig.\ref{fig:arbitrary_rectangle} (c), (d) are generated, which PCMF1 cannot handle accurately. Also, more elongated shapes like (e) are generated for XCI when the WDM channels are not all identical and uniformly spaced.

However, In PCFM2, any shape could be covered using a few adjacent rectangles, to any desired accuracy (see Fig.\ref{fig:arbitrary_rectangle}, red dashed lines). Accordingly, the NLI PSD is derived in closed form for arbitrary rectangular sub-domains $x$ defined by $a \le f_2 \le b, c \le f_1 \le d$, as shown in Fig.~\ref{fig:arbitrary_rectangle} (f):
\begin{equation}
    \begin{aligned}
      G_{\text{NLI},x}\left( f \right) & =\frac{16}{27}{{\Gamma }}\left( f \right) G_{\mathrm{WDM},m_{\mathrm{ch}}}
      G_{\mathrm{WDM},k_{\mathrm{ch}}}G_{\mathrm{WDM},n_{\mathrm{ch}}}\\
      \int_{a }^{b }{\int_{c }^{d }} 
      & {{\left( {{\gamma_x }}\left( {{f}_{1}},{{f}_{2}},f, z \right) \right)}^{2}} \cdot {{\left| \rho_x  \left( {{f}_{1}},{{f}_{2}},f,z \right) \right|}^{2}}d{{f}_{1}}d{{f}_{2}} \\ 
    \end{aligned}
    \label{eq:GNLI_x}
\end{equation}
where the indices $m_{\mathrm{ch}}$, $k_{\mathrm{ch}}$, and $n_{\mathrm{ch}}$ identify the three WDM channels beating together in the rectangle, producing NLI at frequency $f$ in CUT. The WDM spectra are assumed to be rectangles. The key is reduced to express the double integral in Eq.~(\ref{eq:GNLI_x}) in closed-form:
\begin{equation}
    \begin{aligned}
      K_{x}\left( f \right) & = \int_{a }^{b }{\int_{c }^{d }} 
      {{{{\gamma_x }}^{2}\left( {{f}_{1}},{{f}_{2}},f, z \right) }} \\
       &\cdot {{\left| \rho_x  \left( {{f}_{1}},{{f}_{2}},f,z \right) \right|}^{2}}d{{f}_{1}}d{{f}_{2}} \\ 
    \end{aligned}
    \label{eq:K_x}
\end{equation}
We assume that the link consists of $N_s$ spans, with span lengths \(l_1,l_2,\dots,l_{N_s}\). We define the accumulated span-end coordinates as
\begin{equation}
L_0=0,
\qquad
L_n = \sum_{m=1}^{n} l_m,
\qquad n=1,\dots,N_s .\nonumber
\label{eq:Ln_def}
\end{equation}

The $n\text{-th}$ span is defined as in Fig.~1 in {\cite{2025_JLT_Poggiolini}}, which may differ from the others. The span fiber is characterized through its dispersion $\beta^{(n)}(f)$, total attenuation $\alpha_{T}^{(n)}(f)$ and non-linearity coefficient $\gamma^{(n)}({\bar{f} })$. The complex propagation constant  ${{\kappa }^{(n)}}\left( f,z \right)$ defined as:
\begin{equation}
    \begin{aligned}
    {{\kappa }^{(n)}}\left( f,z \right)=-j{{\beta }^{(n)}}\left( f \right)-{{\alpha_T }^{(n)}}\left( f,z \right)
    \end{aligned}
\end{equation}

More details can be found in {\cite{2025_JLT_Poggiolini}}. Therefore, Eq.~(\ref{eq:K_x}) can be re-written as:

\begin{equation}
    \begin{aligned}
      K_{x}\left( f \right) & = {{e}^{-2\int_{0}^{L_{\text{tot}}}{{{\alpha_T }}\left( f,z \right)dz}}}  \int_{a }^{b }{\int_{c }^{d }} d{{f}_{1}}d{{f}_{2}} 
      \\
      &{{\left| \sum_{n=1}^{N_s}\gamma_x^{(n)}{{\int_{L_{n-1}}^{L_n}{{{e}^{\int_{0}^{z}{\Delta {{\kappa^{(n)}}}\left( {{f}_{1}},{{f}_{2}},f,z' \right) d{z}'}}}}dz }}\ \right|}^{2}}\\ 
    \end{aligned}
    \label{eq:K_x_I}
\end{equation}
where $\gamma_x^{(n)}$ is piecewise constant over each span $n$ for the specific island $x$, and is defined in Eq.(A.7) in \cite{2025_JLT_Poggiolini}. Following a similar procedure discussed in Appendix {A, Sect. (B)} in \cite{2025_JLT_Poggiolini}, we obtain:
\begin{equation}
\begin{aligned}
K_{x}(f) =& p(f,L_{\mathrm{tot}})
\int_{a}^{b} \int_{c}^{d}\Bigg|
\sum_{n=1}^{N_s}\int_{L_{n-1}}^{L_n} \gamma_x^{(n)}\cdot p_x^{(n)}(z) \\
& \cdot
\exp\Bigg(
j4\pi^{2} f_{1}f_{2}C(z)
\Bigg)
\, dz
\Bigg|^{2}df_{1}\,df_{2} \\
C(z) =&
\sum_{m=1}^{n-1} \beta_{2,\mathrm{eff},x}^{(m)} l_m
+ \beta_{2,\mathrm{eff},x}^{(n)} \left(z-L_{n-1}\right)
\end{aligned}
\label{eq:K_x_II}
\end{equation}
with the effective SPP in the $n\text{-th}$ span as:
\begin{equation}
    \begin{aligned}           
    p_x^{(n)}(z) =  \sqrt{\frac{
    p_{m_{\text{ch}}}^{(n)}(z)\,
    p_{k_{\text{ch}}}^{(n)}(z)\,
    p_{n_{\text{ch}}}^{(n)}(z)
    }{
    p_{\text{CUT}}^{(n)}(z)
    }}
    \label{eq:rho_general}
    \end{aligned}
\end{equation}
The effective dispersion $\beta_{2,\mathrm{eff},x}^{(n)}$ is piecewise constant over each span $n$ for the specific island $x$.
$p(f,L_{\mathrm{tot}})$ denotes the accumulated gain/loss factor at frequency $f$ along the whole link as:
\begin{equation}
p(f,L_{\mathrm{tot}})
=
e^
{-2\int_{0}^{L_{\mathrm{tot}}}\alpha_T(f,\zeta)\,d\zeta}
\label{eq:p_f_Ltot}
\end{equation}

Then we introduce the local span coordinate
\(\xi=z-L_{n-1}\), and the accumulated dispersion satisfies: 
\begin{equation}
C(L_{n-1}+\xi)
=
\sum_{m=1}^{n-1}\beta_{2,\mathrm{eff},x}^{(m)}l_m
+
\beta_{2,\mathrm{eff},x}^{(n)}\xi ,
\qquad 0\le \xi\le l_n.\nonumber
\label{eq:Cv_local}
\end{equation}

By defining the inter integral over $z$ in the $n\text{-th}$ span as:
\begin{equation}
\begin{aligned}
A_n(f_1,f_2)
&=\gamma_x^{(n)}
\int_0^{l_n}
p_x^{(n)}(\xi)\,
e^{j4\pi^2 f_1 f_2 C(L_{n-1}+\xi)}
\,d\xi .
\label{eq:An_def}
\end{aligned}
\end{equation}
   
\begin{equation}
\begin{aligned}
K_{x}(f) & = p(f,L_{\mathrm{tot}})
\sum_{v=1}^{N_s}\sum_{w=1}^{N_s} K_x^{(v,w)}(f)\\
\end{aligned}
\label{eq:sum_vw_before_cos_sum}
\end{equation}
where $K_x^{(v,w)}(f)$ describes the cross-correlation between the $v\text{-th}$  and $w\text{-th}$ spans:
\begin{equation}
\begin{aligned}
K_x^{(v,w)}&(f) =\gamma_x^{(v)}\gamma_x^{(w)}\int_a^b\int_c^d
\int_0^{l_v}\int_0^{l_w}
p_x^{(v)}(\xi)\,p_x^{(w)}(\eta)
\\
&\cdot e^{j4\pi^2 f_1 f_2 \left[C(L_{w-1}+\eta)-C(L_{v-1}+\xi)\right]}
\,d\xi\,d\eta\,df_1\,df_2 .
\end{aligned}
\label{eq:sum_vw_before_cos}
\end{equation}

Swapping $v$ and $w$ in Eq.~(\ref{eq:sum_vw_before_cos}) changes only the exponential to its complex conjugate, we thus obtain:
\begin{equation}
\begin{aligned}
&\mathcal{R}_x^{(v,w)}(f)
=K_x^{(v,w)}(f) + K_x^{(w,v)}(f)\\
&=2\gamma_x^{(v)}\gamma_x^{(w)}\int_{a }^{b }\int_{c }^{d }
\int_0^{l_v}\int_0^{l_w} p_x^{(v)}(\xi)\,p_x^{(w)}(\eta) \\
&
\cos\!\Big(
4\pi^2 f_1 f_2 [C(L_{w-1}+\eta)-C(L_{v-1}+\xi)]
\Big)
\,d\xi\,d\eta\,df_1\,df_2
\end{aligned}
\label{eq:Kvw_coh_general}
\end{equation}

When $v=w$, all variables are local. It reduces to auto-correlation, representing the intra-span coherence:
\begin{equation}
\begin{aligned}
K_x^{(v,v)}(f) & = \left(\gamma_x^{(v)}\right)^2
\int_a^b\int_c^d
\int_0^{l_v}\int_0^{l_v}
p_x^{(v)}(\xi)\,p_x^{(v)}(\eta)
\\
&\cdot e^{j4\pi^2 f_1 f_2 \beta_{2,\mathrm{eff},x}^{(v)} (\eta-\xi)}
\,d\xi\,d\eta\,df_1\,df_2 \\
&=\frac{1}{2}\mathcal{R}_x^{(v,v)}(f)
\end{aligned}
\label{eq:Kvv_def_expanded}
\end{equation}

Therefore, \(K_{x}(f)\) can be re-written as
\begin{equation}
K_{x}(f) =
p(f,L_{\mathrm{tot}})
\left[\frac{1}{2}
\sum_{v=1}^{N_s}\mathcal{R}_x^{(v,v)}(f)
+
\sum_{v=1}^{N_s}\sum_{w=1}^{v-1}
\mathcal{R}_x^{(v,w)}(f)
\right]
\label{eq:K_diag_pair}
\end{equation}

Up to this point, the key extensions of PCFM2 relative to PCFM1 have been established. The formulation is developed at an arbitrary frequency \(f\), rather than being restricted to a locally white approximation, and the NLI coherent accumulation is explicitly retained instead of being neglected through incoherent accumulation.

Moreover, by representing each original nonlinear interaction island as a union of rectangular sub-domains, the framework is no longer restricted to approximated SCI/XCI geometries and can be naturally extended to arbitrary integration regions.

Therefore, the proposed model removes the limitations associated with assumptions i)-iv) and provides a unified basis for the closed-form development that follows. The main objective of this closed-form evaluation is the generalized kernel \(\mathcal{R}_x^{(v,w)}(f)\), that will be addressed in Sect.~(\ref{sect:PCFM2}).

\section{PCFM2 expression}
\label{sect:PCFM2}
To obtain a closed-form expression, we express the effective SPP in Eq.~(\ref{eq:rho_general}) by a $N_p$-degree polynomial: 
\begin{equation}
p_x^{(v)}(\xi)=\sum_{n_1=0}^{N_p} a_{v,n_1}\,\xi^{n_1},
\quad
p_x^{(w)}(\eta)=\sum_{n_2=0}^{N_p} a_{w,n_2}\,\eta^{n_2}
\end{equation}
Substituting into Eq.~(\ref{eq:Kvw_coh_general}) yields:
\begin{equation}
\begin{aligned}
&\mathcal{R}_x^{(v,w)}(f)=2\gamma_x^{(v)}\gamma_x^{(w)}
\sum_{n_1=0}^{N_p}\sum_{n_2=0}^{N_p}
a_{v,n_1}a_{w,n_2}\,I_{n_1,n_2}^{(v,w)}
\end{aligned}
\label{eq:Kvw_coh_general_II}
\end{equation}
 with:
\begin{equation}
\begin{aligned}
I_{n_1,n_2}^{(v,w)}
&= \int_0^{l_v}\int_0^{l_w}
\xi^{n_1}\eta^{n_2}\\
&F_{a,b;c,d}\!\left(C_{vw} - \beta_{2,\mathrm{eff},v}\xi
+\beta_{2,\mathrm{eff},w}\eta\right)\,d\eta\,d\xi,
\\
F_{a,b;c,d}(t)&=\int_a^b \int_c^d\cos\!\Big(4\pi^2 f_1 f_2 t\Big)
\,df_1\,df_2
\end{aligned}
\label{eq:Cnm_original}
\end{equation}
For convenience, we define
\[
C_{vw}=C(L_{w-1})-C(L_{v-1}).
\]
which is z-independent.
The frequency integral can be evaluated in closed form as shown in Appendix~\ref{app:Mr}:
\begin{equation}
F_{a,b;c,d}(t)
=
\frac{1}{4\pi^2 t}
\sum_{\ell=1}^4 \sigma_\ell\,\text{SI}(\lambda_\ell t),
\label{eq:F_abcd}
\end{equation}
with
\begin{equation}
\lambda_1=4\pi^2 ad,\quad
\lambda_2=4\pi^2 ac,\quad
\lambda_3=4\pi^2 bc,\quad
\lambda_4=4\pi^2 bd,\nonumber
\end{equation}
and
\begin{equation}
(\sigma_1,\sigma_2,\sigma_3,\sigma_4)=(-1,+1,-1,+1).\nonumber
\end{equation}
Note that $F_{a,b;c,d}(0)
=
(b-a)(d-c)$, and the apparent singularity in Eq.~(\ref{eq:F_abcd}) is removable.
For ease of notation, in the remainder of this section
we denote \(\beta_{2,\mathrm{eff},v}\) and \(\beta_{2,\mathrm{eff},w}\)
by \(\beta_v\) and \(\beta_w\), respectively. Introducing the scaled variables
\begin{equation}
x=\beta_v\xi, \qquad y=\beta_w\eta,\nonumber
\end{equation}
and
\begin{equation}
\Delta_v=\beta_v l_v, \qquad \Delta_w=\beta_w l_w,\nonumber
\end{equation}
Eq.~(\ref{eq:Cnm_original}) becomes
\begin{equation}
\begin{aligned}
I_{{n_1,n_2}}^{(v,w)}
&=
\frac{1}{\beta_v^{{n_1}+1}\beta_w^{{n_2}+1}}
\int_0^{\Delta_v}\int_0^{\Delta_w}
x^{n_1} y^{n_2}\\
&F_{a,b;c,d}(y-x+C_{vw})\,dy\,dx  
\end{aligned}
\end{equation}
For compactness, the following derivation is written for
\(\beta_v>0\) and \(\beta_w>0\), or equivalently
\(\Delta_v>0\) and \(\Delta_w>0\), so that the transformed
integration intervals are positively oriented. If signed
dispersion values are retained, the transformed interval
endpoints are first ordered, and the same piecewise-polynomial
construction is then applied. The derivation below also assumes $\beta_v \beta_w \neq 0$. 
The zero-dispersion limit is obtained directly from Eq.~(\ref{eq:Cnm_original}) by setting the corresponding phase term to zero. 
In implementation, this limiting expression is used whenever $|\beta_v|$ or $|\beta_w|$ 
falls below a prescribed numerical threshold. 

We then introduce the key variable $s = y - x + C_{vw}$, which maps the rectangular domain into the interval:
\begin{equation}
s \in [C_{vw}-\Delta_v,\,C_{vw}+\Delta_w]
\end{equation}

The integral can then be reduced to
\begin{equation}
I_{{n_1,n_2}}^{(v,w)}
=
\frac{1}{\beta_v^{n_1+1}\beta_w^{n_2+1}}
\int_{C_{vw}-\Delta_v}^{C_{vw}+\Delta_w}
Q_{n_1,n_2}(s)\,F_{a,b;c,d}(s)\,ds,
\label{eq:Cnm_reduced}
\end{equation}
where
\begin{equation}
Q_{n_1,n_2}(s)
=\int_{x_L(s)}^{x_U(s)}
x^{n_1} (s-C_{vw}+x)^{n_2}\,dx.
\label{eq:Qnm_integral}
\end{equation}
with $x_L(s)=\max(0,\,C_{vw}-s),
x_U(s)=\min(\Delta_v,\,C_{vw}+\Delta_w-s)$. Expanding the integrand in Eq.~(\ref{eq:Qnm_integral}) yields:
\begin{equation}
\begin{aligned}
Q_{n_1,n_2}(s)
&=
\sum_{k=0}^{n_2}
\binom{n_2}{k}
\frac{(s-C_{vw})^{n_2-k}}{n_1+k+1}
\\
&\left[
x_U(s)^{n_1+k+1}-x_L(s)^{n_1+k+1}
\right].    
\end{aligned}
\label{eq:Qnm_general}
\end{equation}

The functions $x_L(s)$ and $x_U(s)$ induce a piecewise structure over the interval $s \in [C_{vw}-\Delta_v,\,C_{vw}+\Delta_w]$,
with breakpoints
\begin{equation}
\begin{aligned}
&u_0=C_{vw}-\Delta_v,\quad
u_1=C_{vw},\quad
\\
&u_2=C_{vw}+\Delta_w-\Delta_v,\quad
u_3=C_{vw}+\Delta_w.    
\end{aligned}
\end{equation}

Depending on the relative magnitude of $\Delta_v$ and $\Delta_w$, the ordering of $(u_1,u_2)$ changes, leading to two configurations. In both cases, $Q_{n_1,n_2}(s)$ consists of three polynomial segments $Q_{n_1,n_2}^{(j)}(s)$ obtained by substituting the corresponding expressions of $x_L(s)$ and $x_U(s)$ into Eq.~(\ref{eq:Qnm_general}).

On each segment $j$, $Q_{n_1,n_2}^{(j)}(s)$ is a polynomial of degree at most $n_1+n_2+1$:
\begin{equation}
Q_{n_1,n_2}^{(j)}(s)=\sum_{r=0}^{n_1+n_2+1} A_r^{(j)} s^r.
\label{eq:Qnm_j}
\end{equation}
The explicit piecewise forms of $Q_{n_1,n_2}(s)$ and the corresponding
coefficients $A_r^{(j)}$ are given in Appendix~\ref{app:Qnm}.

Substituting Eq.~(\ref{eq:Qnm_j}) into Eq.~(\ref{eq:Cnm_reduced}), we obtain
\begin{equation}
\begin{aligned}
I_{n_1,n_2}^{(v,w)}
&=
\frac{1}{\beta_v^{n_1+1}\beta_w^{n_2+1}}
\sum_{j=0}^{2}
\int_{s_j}^{s_{j+1}}
Q_{n_1,n_2}^{(j)}(s)\,F_{a,b;c,d}(s)\,ds
\\
&=
\frac{1}{\beta_v^{n_1+1}\beta_w^{n_2+1}}
\sum_{j=0}^{2}
\sum_{r=0}^{n_1+n_2+1}
A_r^{(j)}\\
&\Bigg[
\int_{0}^{s_{j+1}} s^r F_{a,b;c,d}(s)\,ds
-\int_{0}^{s_{j}} s^r F_{a,b;c,d}(s)\,ds
\Bigg]
\end{aligned}
\label{eq:Inm_piecewise}
\end{equation}

Defining the primitive functions
\begin{equation}
M_r(x)=\int_0^x s^r F_{a,b;c,d}(s)\,ds,
\end{equation}
and 
we obtain
\begin{equation}
M_r(x)
=
\frac{1}{4\pi^2}
\sum_{\ell=1}^4
\sigma_\ell\,H_r^{(\lambda_\ell)}(x),
\label{eq:Mr_expanded}
\end{equation}
where
\begin{equation}
H_r^{(\lambda_\ell)}(x)=\int_0^x s^{r-1} \text{SI}(\lambda_\ell s)\,ds.
\end{equation}

Substituting $M_r(x)$ into Eq.~(\ref{eq:Inm_piecewise}), we obtain
\begin{equation}
I_{n_1,n_2}^{(v,w)}
=
\frac{1}{\beta_v^{n_1+1}\beta_w^{n_2+1}}
\sum_{j=0}^{2}
\sum_{r=0}^{n_1+n_2+1}
A_r^{(j)}
\left[
M_r(s_{j+1})-M_r(s_j)
\right].
\label{eq:Cnm_final}
\end{equation}
The explicit expressions of the coefficients $A_r^{(j)}$ are reported in Appendix~\ref{app:Qnm}. The primitive functions $M_r(x)$ is defined for both positive and negative $x$, with the usual oriented-integral convention. It can be evaluated in closed form in terms of sine-integral functions and finite trigonometric expansions in Appendix~\ref{app:Mr}. Therefore, Eq.~(\ref{eq:Cnm_final}) provides a complete closed-form representation of the coherence kernels for arbitrary span pairs and arbitrary frequency rectangles.

It is worth emphasizing that the same formulation also applies to the
intra-span case by setting $v=w$, so that intra-span coherence is
naturally recovered as a special case of the general pairwise framework. The detailed derivation
is provided in Appendix~\ref{app:single_span_equivalence}.

\section{Validation}
\begin{figure}
    \centering
    \includegraphics[width=0.68\columnwidth]{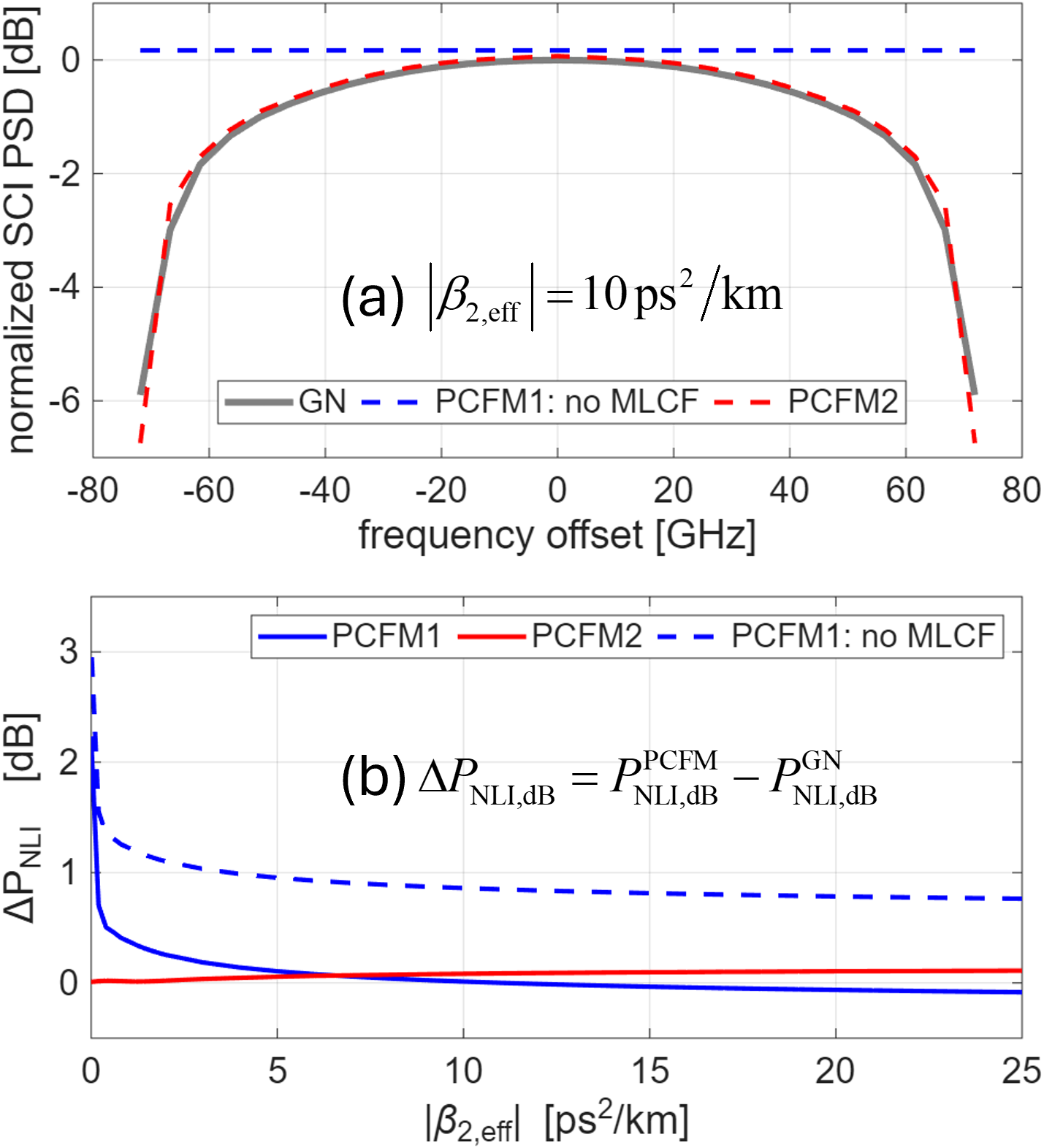}
    \caption{(a) Normalized SCI PSD in a single 140~GBaud channel after 100~km. 
    (b) SCI power error for PCFM1 and PCFM2, with respect to the GN model over $|\beta_{2,\mathrm{eff}}|\in[0,25]~\mathrm{ps}^2/\mathrm{km}$.}
    \label{fig:SCI_PSD}
\end{figure}

\begin{figure}
    \centering
    \includegraphics[width=0.68\columnwidth]{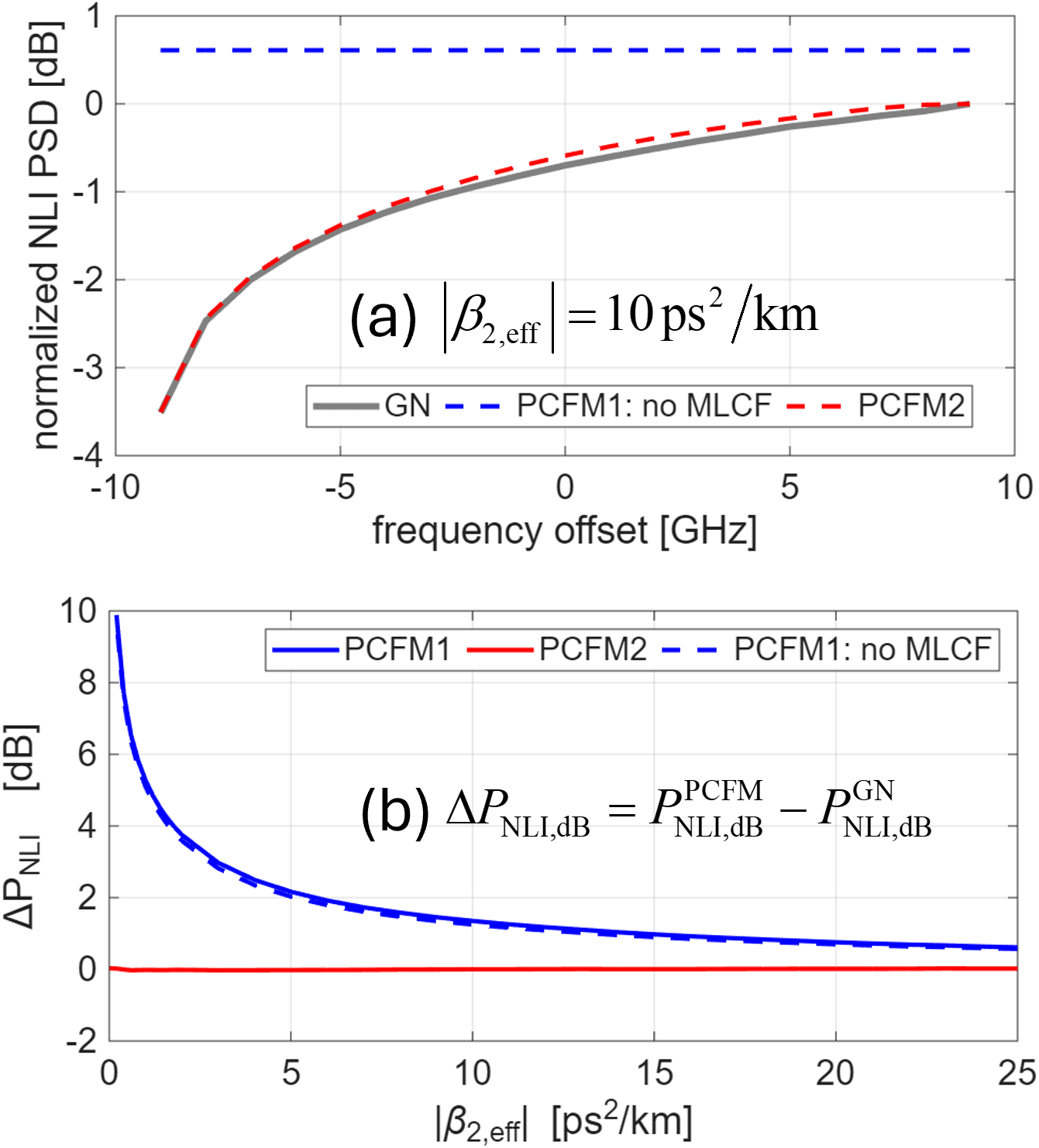}
    \caption{(a) Normalized NLI PSD in the first subcarrier after 100~km, (b) NLI power error of the first subcarrier in an eight-subcarrier system for PCFM1 and PCFM2, with respect to the GN model over $|\beta_{2,\mathrm{eff}}|\in[0,25]~\mathrm{ps}^2/\mathrm{km}$.}
    \label{fig:NLI_subcarrier}
\end{figure}

\begin{figure}
    \centering
    \includegraphics[width=0.8\linewidth]{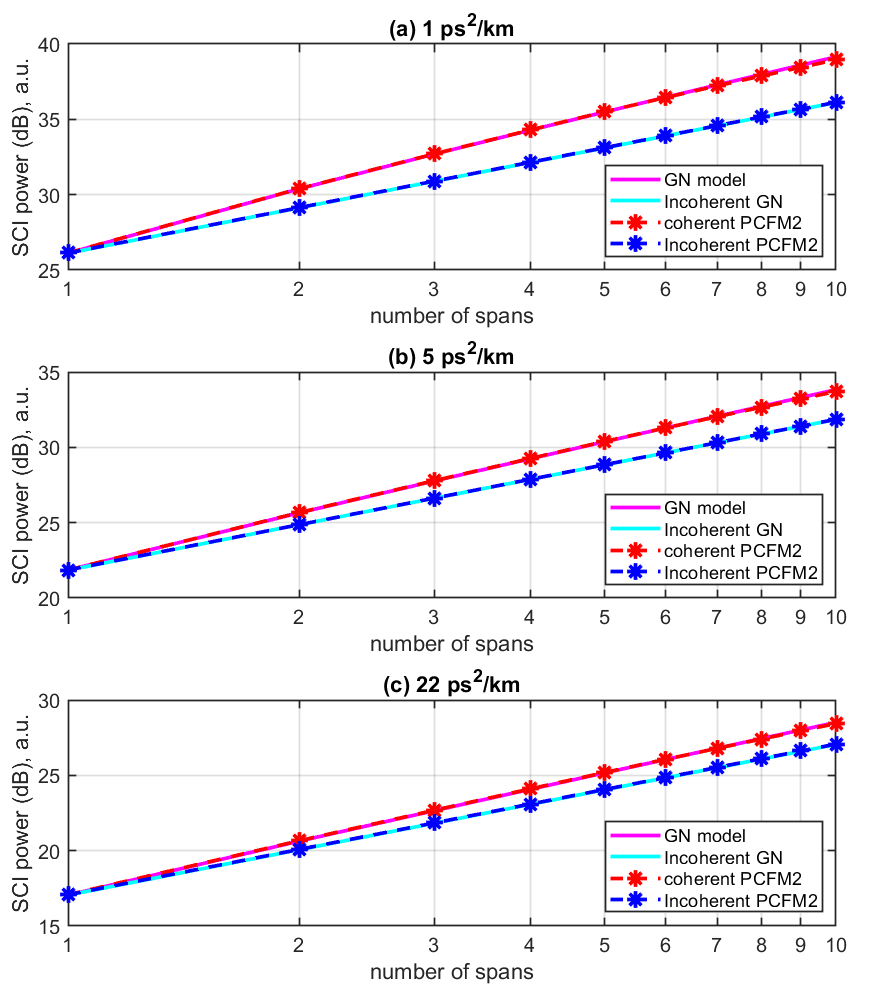}
    \caption{Single-channel 140 GBaud, with different  dispersion values: (a) $|\beta_{2,\mathrm{eff}}| = 1$ ps$^2$/km, 
(b) $|\beta_{2,\mathrm{eff}}| = 5$ ps$^2$/km, 
(c) $|\beta_{2,\mathrm{eff}}| = 22$ ps$^2$/km. Cyan and blue: incoherent NLI accumulation.
    Magenta and red: coherent NLI accumulation. Solid are reference GN, Dashed with markers PCFM2.}
    \label{fig:coherent}
\end{figure}
In this section, we validate the proposed PCFM2 formulation through three representative tests. Note that MLCF was not used, since it is no longer needed.

\subsection*{Test 1: the SCI power spectral density (PSD)}
This test validates the removal of the locally-white NLI approximation and the ability to evaluate the NLI contribution at arbitrary frequency offsets within the CUT. We focus on the computation of the accurate full PSD of SCI of a 140~GBaud channel transmitted over 100~km. In this test, the channel SPP is that of the 100th channel of the C+L+S system of \cite{2025_JLT_Poggiolini}(Fig.~11), representative of strong ISRS and backward Raman amplification. We assume $N_p=5$. 
The reference GN-model result (through numerical integration) is the gray solid line in Fig.~\ref{fig:SCI_PSD}(a). PCFM1 returns a straight line since it can only compute the value at $f$=0, and then assumes a flat (``white'') PSD. PCFM2, instead, can compute the PSD with high accuracy at any frequency.

Figure~\ref{fig:SCI_PSD}(b) reports the resulting SCI overall power error compared to the reference GN model, after DSP matched filtering, over the range $|\beta_{2,\mathrm{eff}}|\in[0,25]~\mathrm{ps}^2/\mathrm{km}$.
PCFM1 (dashed blue line) shows a large error vs. the reference, on the order of 1dB. When the MLCF is used, such error is mitigated (solid blue line). However, for low dispersion the MLCF is ineffective and the error goes back up steeply. PCFM2 is instead very accurate across all dispersions, including arbitrarily close to zero. Such accuracy is not obtained through the complex training of a MLCF, but thanks to the intrinsically accurate PCFM2 formula.

\subsection*{Test 2: multi-subcarrier systems}
This test validates the rectangular-domain formulation for arbitrary islands, including MCI terms that are neglected in PCFM1. We consider a subcarrier-based channel. It is made up of 8 identical subcarriers, each operating at 18~GBaud with rectangular spectra and, initially, with no guard band. We focus on the leftmost subcarrier and compare the NLI PSD predicted by the reference GN model, PCFM1, and PCFM2. Note that the NLI PSD here is not due to just SCI, but also includes XCI and MCI from the other 7 subcarriers. Fig.~\ref{fig:NLI_subcarrier}(a) shows the obtained PSD. In this example the deviation from flat PSD is quite substantial and asymmetric, and PCFM2 captures it accurately. Fig.~\ref{fig:NLI_subcarrier}(b) shows the total NLI power error compared to the reference GN model vs. dispersion, showing a remarkably low error for PCFM2, whereas the error is quite significant for PCFM1. Here the MLCF is not effective in improving PCFM1: whether it would be possible to train it for low-dispersion multi-subcarrier scenarios is an open issue, but PCFM2 makes it irrelevant.

\subsection*{Test 3: multispan coherence}
To validate the accuracy of the proposed multispan coherence formulation, we considere a single-channel transmission system. The channel symbol rate was set to 140 GBaud. The transmission link consists of 10 spans, 100 km each. To investigate the impact of dispersion on coherent NLI accumulation, three values of effective dispersion are considered, namely 
$|\beta_{2,\mathrm{eff}}| = 1$, $5$, and $22~\mathrm{ps}^2/\mathrm{km}$, and
kept constant along the link in each case. Each span was assumed to be completely compensated for using only backward Raman amplification, resulting in a significant power increase toward the end of the fiber and consequently enhance inter-span coherence effect.

The reference results were obtained from both the incoherent and coherent GN-model along the link. For comparison, we also provided the incoherent and coherent PCFM2, as shown in Fig.~\ref{fig:coherent}. When coherence is taken into account, the SCI power increases faster with the number of spans than in the incoherent case. This is expected, since the phase correlation among the NLI contributions generated in different spans leads to an additional constructive accumulation. As a result, the gap between coherent and incoherent accumulation becomes progressively larger as the link length increases. 

Moreover, by comparing different dispersion regimes, it can be observed that the impact of coherence becomes more pronounced as the dispersion decreases. In the high-dispersion case (e.g., $|\beta_{2,\mathrm{eff}}| = 22~\mathrm{ps}^2/\mathrm{km}$), the coherent and incoherent curves remain relatively close, indicating that phase decorrelation across spans partially suppresses coherent accumulation. In contrast, for lower dispersion (e.g., $|\beta_{2,\mathrm{eff}}| = 5~\mathrm{ps}^2/\mathrm{km}$), the gap between coherent and incoherent accumulation becomes significantly larger, showing that phase correlation is better preserved and coherence effects are stronger. 

The coherent PCFM2 curve closely follows the coherent GN reference over all considered spans and dispersion values. This demonstrates that the proposed analytical formulation is able to accurately capture inter-span coherence effects. In particular, the agreement remains very good even at larger span counts, where coherence effects become more evident.

\section{Conclusion}
We have presented the PCFM2 that removes key approximations commonly adopted in existing closed-form formulations, including spectrally flat PSD, simplified integration domains, neglect of MCI, and incoherent accumulation across spans. By introducing an accumulated-dispersion-based formulation and leveraging polynomial SPP representation, the proposed model enables a fully closed-form evaluation of both intra-span and inter-span NLI contributions. The validation against numerical GN-model integration confirms that the proposed framework accurately evaluates the spectral NLI PSD, accounts for SCI/XCI/MCI contributions in multi-subcarrier systems, and captures coherent NLI accumulation across spans. These results demonstrate the relevance of PCFM2 for low-dispersion and digital-subcarrier scenarios, while retaining the computational efficiency required for practical system analysis and optimization.

\begin{appendices}

\section{Explicit Expressions of the Overlap Polynomial}
\label{app:Qnm}

In this appendix, to simplify the notation, we denote the polynomial indices
\(n_1\) and \(n_2\) used in the main text by \(n\) and \(m\), respectively.
These indices should not be confused with span or segment indices used elsewhere. Here We provide the explicit expressions of the overlap polynomial $Q_{nm}(s)$ and its coefficients.

Recall that
\begin{equation}
Q_{n,m}(s)
=\int_{x_L(s)}^{x_U(s)}
x^{n} (s-C_{vw}+x)^{m}\,dx,
\end{equation}
where
\begin{equation}
x_L(s)=\max(0,\,C_{vw}-s),
x_U(s)=\min(\Delta_v,\,C_{vw}+\Delta_w-s).
\end{equation}

Expanding the integrand yields
\begin{equation}
(s-C_{vw}+x)^m
=
\sum_{k=0}^m
\binom{m}{k}(s-C_{vw})^{m-k}x^k,
\end{equation}
and therefore
\begin{equation}
\begin{aligned}
Q_{nm}(s)
&=
\sum_{k=0}^m
\binom{m}{k}
\frac{(s-C_{vw})^{m-k}}{n+k+1}
\\
&\left[
x_U(s)^{n+k+1}-x_L(s)^{n+k+1}
\right].    
\end{aligned}
\label{eq:Qnm_general_appendix}
\end{equation}

The admissible range of $s$ is
\[
s\in [\,C_{vw}-\Delta_v,\;C_{vw}+\Delta_w\,],
\]
which follows from
\[
s=y-x+C_{vw},
\qquad
0\le x\le \Delta_v,\quad 0\le y\le \Delta_w.
\]
The explicit coefficients $A_r^{(j)}$ below are written for
\(\Delta_v>0\) and \(\Delta_w>0\). For signed-dispersion
cases, the transformed interval endpoints are first ordered,
and the same piecewise-polynomial construction is then
applied. Define the breakpoints
\begin{equation}
\begin{aligned}
&u_0=C_{vw}-\Delta_v,\quad
u_1=C_{vw},\quad
\\
&u_2=C_{vw}+\Delta_w-\Delta_v,\quad
u_3=C_{vw}+\Delta_w.    
\end{aligned}
\end{equation}
The breakpoints are determined by the values of $s$ at which the
endpoint functions
\[
x_L(s)=\max(0,\,C_{vw}-s),
x_U(s)=\min(\Delta_v,\,C_{vw}+\Delta_w-s)
\]
In particular, $u_1=C_{vw}$ is obtained from
$C_{vw}-s=0$, where $x_L(s)$ switches from $C_{vw}-s$ to $0$, while
$u_2=C_{vw}+\Delta_w-\Delta_v$ is obtained from
$\Delta_v=C_{vw}+\Delta_w-s$, where $x_U(s)$ switches from $\Delta_v$
to $C_{vw}+\Delta_w-s$. The quantities
$u_0=C_{vw}-\Delta_v$ and $u_3=C_{vw}+\Delta_w$ are instead the lower
and upper endpoints of the admissible interval of $s$.

Depending on the relative magnitude of $\Delta_v$ and $\Delta_w$, the ordering of $(u_1,u_2)$ changes, which is divided into two cases:

\subsubsection*{Case A: $\Delta_v\le\Delta_w$}

\begin{equation}
u_0 \le u_1 \le u_2 \le u_3.\nonumber
\end{equation}

\paragraph{Segment 1: $u_0 \le s \le u_1$}
\begin{equation}
x_L=C_{vw}-s,\quad x_U=\Delta_v.\nonumber
\end{equation}

\paragraph{Segment 2: $u_1 \le s \le u_2$}
\begin{equation}
x_L=0,\quad x_U=\Delta_v.\nonumber
\end{equation}

\paragraph{Segment 3: $u_2 \le s \le u_3$}
\begin{equation}
x_L=0,\quad x_U=C_{vw}+\Delta_w-s.\nonumber
\end{equation}

\subsubsection*{Case B: $\Delta_v\ge\Delta_w$}

\begin{equation}
u_0 \le u_2 \le u_1 \le u_3.\nonumber
\end{equation}

\paragraph{Segment 1: $u_0 \le s \le u_2$}
\begin{equation}
x_L=C_{vw}-s,\quad x_U=\Delta_v.\nonumber
\end{equation}

\paragraph{Segment 2: $u_2 \le s \le u_1$}
\begin{equation}
x_L=C_{vw}-s,\quad x_U=C_{vw}+\Delta_w-s.\nonumber
\end{equation}

\paragraph{Segment 3: $u_1 \le s \le u_3$}
\begin{equation}
x_L=0,\quad x_U=C_{vw}+\Delta_w-s.\nonumber
\end{equation}

In both cases, $Q_{nm}(s)$ is represented by three segment-wise
polynomial branches, denoted by $Q_{nm}^{(j)}(s)$.On each segment, the functions $x_L(s)$ and $x_U(s)$ reduce to either
constants or affine functions of $s$. Therefore,
$x_U(s)^{n+k+1}$ and $x_L(s)^{n+k+1}$ are polynomials in $s$ of degree
at most $n+k+1$, while $(s-C_{vw})^{m-k}$ has degree $m-k$.
Hence each term $Q_{nm}^{(j)}(s)$ has degree at most $(m-k)+(n+k+1)=n+m+1$:
\begin{equation}
Q_{nm}^{(j)}(s)=\sum_{r=0}^{n+m+1} A_r^{(j)} s^r.
\end{equation}

The coefficients $A_r^{(j)}$ are obtained by expanding Eq.~(\ref{eq:Qnm_general_appendix}) and collecting powers of $s$.

We adopt the convention
\begin{equation}
\binom{p}{q}=0 \quad \text{if } q<0 \text{ or } q>p. \nonumber
\end{equation}

Segment 1 coefficients can be written as:
\begin{equation}
\begin{aligned}
&A_r^{(1)}
=
\sum_{k=0}^m
\binom{m}{k}
\frac{\Delta_v^{\,n+k+1}}{n+k+1}
\binom{m-k}{r}
(-C_{vw})^{m-k-r}
\\
&-
\left[
\sum_{k=0}^m
\binom{m}{k}
\frac{(-1)^{n+k+1}}{n+k+1}
\right]
\binom{n+m+1}{r}
(-C_{vw})^{n+m+1-r}.    
\end{aligned}
\end{equation}

Segment 2 coefficients (Case A) can be written as:
\begin{equation}
A_r^{(2A)}
=
\sum_{k=0}^m
\binom{m}{k}
\frac{\Delta_v^{n+k+1}}{n+k+1}
\binom{m-k}{r}
(-C_{vw})^{m-k-r}.
\end{equation}

Segment 2 coefficients (Case B) can be written as:
\begin{equation}
\begin{aligned}
&A_r^{(2B)}
=
\sum_{k=0}^m
\binom{m}{k}
\frac{1}{n+k+1}
\sum_{q=0}^{m-k}
\binom{m-k}{q}
\binom{n+k+1}{r-q}
\\
&\cdot
(-C_{vw})^{m-k-q}
(-1)^{r-q}
(C_{vw}+\Delta_w)^{n+k+1-(r-q)}
\\
&
-
\left[
\sum_{k=0}^m
\binom{m}{k}
\frac{(-1)^{n+k+1}}{n+k+1}
\right]
\binom{n+m+1}{r}
(-C_{vw})^{n+m+1-r}.
\end{aligned}    
\end{equation}

Segment 3 coefficients can be written as:
\begin{equation}
\begin{aligned}
A_r^{(3)}
&=
\sum_{k=0}^m
\binom{m}{k}
\frac{1}{n+k+1}
\sum_{q=0}^{m-k}
\binom{m-k}{q}
\binom{n+k+1}{r-q}
\\
&\cdot
(-C_{vw})^{m-k-q}
(-1)^{r-q}
(C_{vw}+\Delta_w)^{n+k+1-(r-q)}.    
\end{aligned}
\end{equation}

\section{Closed-Form Primitive Functions}
\label{app:Mr}

In this appendix, we provide the closed-form expressions of the primitive functions
\begin{equation}
\begin{aligned}
M_r(x)&=\int_0^x s^r F_{a,b;c,d}(s)\,ds,
\\
F_{a,b;c,d}(s)
&=
\int_a^b \int_c^d \cos\!\Big(
4\pi^2 f_1 f_2
\cdot s
\Big)
\,df_1\,df_2.    
\end{aligned}
\end{equation}

Define
\begin{equation}
\lambda_1=4\pi^2 ad,\quad
\lambda_2=4\pi^2 ac,\quad
\lambda_3=4\pi^2 bc,\quad
\lambda_4=4\pi^2 bd,\nonumber
\end{equation}
with signs
\begin{equation}
(\sigma_1,\sigma_2,\sigma_3,\sigma_4)=(-1,+1,-1,+1).\nonumber
\end{equation}

We first derive a closed-form expression for $F_{a,b;c,d}(s)$. Integrating
first with respect to $f_1$, we obtain
\begin{equation}
\begin{aligned}
F_{a,b;c,d}(s)
&=
\int_a^b
\left[
\int_c^d \cos\!\big(4\pi^2 f_1 f_2 s\big)\,df_1
\right]df_2
\\
&=
\int_a^b
\frac{1}{4\pi^2 f_2 s}
\left[
\sin\!\big(4\pi^2 d f_2 s\big)
-
\sin\!\big(4\pi^2 c f_2 s\big)
\right]
df_2
\\
&=
\frac{1}{4\pi^2 s}
\sum_{\ell=1}^4 \sigma_\ell\,SI(\lambda_\ell s),
\end{aligned}
\end{equation}
where $SI$ is the sinint function.
Then we have
\begin{equation}
M_r(x)
=
\frac{1}{4\pi^2}
\sum_{\ell=1}^4
\sigma_\ell\,H_r^{(\lambda_\ell)}(x),
\end{equation}
where
\begin{equation}
H_r^{(\lambda)}(x)=\int_0^x s^{r-1}SI(\lambda s)\,ds.
\end{equation}
The evaluation of $H_r^{(\lambda)}(x)$ depends on whether $r=0$ or $r\ge 1$,
and we treat these two cases separately below.
\subsection{Case $r=0$}

Define
\begin{equation}
J(X)=\int_0^X \frac{SI(t)}{t}\,dt,
\end{equation}
then
\begin{equation}
H_0^{(\lambda)}(x)=J(\lambda x),
\end{equation}
with closed form
\begin{equation}
J(X)
=
X\,{}_2F_3\!\left(
\frac12,\frac12;
\frac32,\frac32,\frac32;
-\frac{X^2}{4}
\right).
\end{equation}

\subsection{Case $r\ge1$}

For $r\ge 1$, we start from
\begin{equation}
H_r^{(\lambda)}(x)=\int_0^x s^{r-1}SI(\lambda s)\,ds.
\end{equation}

We first introduce the change of variable
\begin{equation}
X=\lambda s,
\qquad
s=\frac{X}{\lambda},
\qquad
ds=\frac{dX}{\lambda}.\nonumber
\end{equation}
Then
\begin{equation}
\begin{aligned}
H_r^{(\lambda)}(x)
&=
\int_0^x s^{r-1}SI(\lambda s)\,ds
\\
&=
\int_0^{\lambda x}
\left(\frac{X}{\lambda}\right)^{r-1}
SI(X)\,
\frac{dX}{\lambda}
\\
&=
\frac{1}{\lambda^r}
\int_0^{\lambda x} X^{r-1}SI(X)\,dX.
\end{aligned}
\end{equation}

We now integrate by parts. Let
\begin{equation}
u=SI(X),
\qquad
dv=X^{r-1}dX.\nonumber
\end{equation}
Then
\begin{equation}
du=\frac{\sin X}{X}\,dX,
\qquad
v=\frac{X^r}{r}.\nonumber
\end{equation}
Hence
\begin{equation}
\begin{aligned}
&\int_0^{\lambda x} X^{r-1}SI(X)\,dX
\\
&=
\left[\frac{X^r}{r}SI(X)\right]_0^{\lambda x}
-
\frac{1}{r}\int_0^{\lambda x}X^r\frac{\sin X}{X}\,dX
\\
&=
\left[\frac{X^r}{r}SI(X)\right]_0^{\lambda x}
-
\frac{1}{r}\int_0^{\lambda x}X^{r-1}\sin X\,dX.\nonumber
\end{aligned}
\end{equation}

Define
\begin{equation}
S_n(X)=\int_0^X t^n\sin t\,dt.
\end{equation}
Then
\begin{equation}
\int_0^{\lambda x}X^{r-1}\sin X\,dX=S_{r-1}(\lambda x),
\end{equation}
so that
\begin{equation}
H_r^{(\lambda)}(x)
=
\frac{1}{\lambda^r}
\left[
\frac{X^r}{r}SI(X)-\frac{1}{r}S_{r-1}(X)
\right]_{0}^{\lambda x}.
\end{equation}

Since $r\ge 1$, the lower-end contribution at $X=0$ vanishes, and therefore
\begin{equation}
H_r^{(\lambda)}(x)
=
\frac{1}{\lambda^r}
\left[
\frac{X^r}{r}SI(X)-\frac{1}{r}S_{r-1}(X)
\right]_{X=\lambda x}.
\end{equation}

The function $S_n(X)$ admits the expansion
\begin{equation}
\begin{aligned}
S_n(X)
&=
\sum_{p=0}^{n}
(-1)^p\frac{n!}{(n-p)!}X^{n-p}
\sin\!\left(X-\frac{\pi}{2}(p+1)\right)
\\
&+
n!\sin\!\left(\frac{\pi}{2}(n+1)\right).    
\end{aligned}
\end{equation}

Therefore, the primitive function $M_r(x)$ admits the following fully
closed-form expression:
\begin{equation}
\begin{aligned}
&M_r(x)=
\frac{1}{4\pi^2}
\sum_{\ell=1}^4
\sigma_\ell\,H_r^{(\lambda_\ell)}(x)=
\\
&
\begin{cases}
\frac{1}{4\pi^2}
\sum_{\ell=1}^4
\sigma_\ell\,J(\lambda_\ell x), &r=0,\\[0.8em]
\displaystyle
\frac{1}{4\pi^2}
\sum_{\ell=1}^4
\sigma_\ell\,
\frac{1}{\lambda_\ell^r}
\left[
\frac{(\lambda_\ell x)^r}{r}SI(\lambda_\ell x)-\frac{1}{r}S_{r-1}(\lambda_\ell x)
\right],
&r\ge 1,
\end{cases}
\end{aligned}
\end{equation}
where
\begin{equation}
\lambda_1=4\pi^2 ad,\quad
\lambda_2=4\pi^2 ac,\quad
\lambda_3=4\pi^2 bc,\quad
\lambda_4=4\pi^2 bd,\nonumber
\end{equation}
\begin{equation}
(\sigma_1,\sigma_2,\sigma_3,\sigma_4)=(-1,+1,-1,+1),\nonumber
\end{equation}
\begin{equation}
\begin{aligned}
&J(X)=
X\,{}_2F_3\!\left(
\frac12,\frac12;
\frac32,\frac32,\frac32;
-\frac{X^2}{4}
\right),\nonumber
\\
&S_n(X)
=
\sum_{p=0}^{n}
(-1)^p\frac{n!}{(n-p)!}X^{\,n-p}
\sin\!\left(X-\frac{\pi}{2}(p+1)\right)
\\
&\qquad
+
n!\sin\!\left(\frac{\pi}{2}(n+1)\right).\nonumber
\end{aligned}
\end{equation}

\section{Reduction to the Single-Span Closed Form}
\label{app:single_span_equivalence}

In this appendix, we show that in the special case \(v=w\) the proposed multispan coherence formulation reduces to the conventional single-span GN-model contribution i.e., the building block of the usual incoherent multispan accumulation. This provides an important consistency check.

We consider the single-span self-term, corresponding to \(v=w\). In this case, the kernel depends only on the difference variable \(\eta-\xi\):
\begin{equation}
I_{nm}^{(v,v)}
=
\int_0^{l_v}\int_0^{l_v}
\xi^n\eta^m\,
F_{a,b;c,d}\!\bigl(\beta_v(\eta-\xi)\bigr)\,d\eta\,d\xi,
\ n,m\ge 0.
\label{eq:Cvv_def_app}
\end{equation}

In this appendix, to simplify the notation, we denote the polynomial indices
\(n_1\) and \(n_2\) used in the Eq.~(\ref{eq:Cnm_original}) by \(n\) and \(m\), respectively.
These indices should not be confused with span or segment indices used elsewhere. For the single-span case $C_{vv}=0$ and $\Delta_v=\Delta_w=\Delta=\beta_vL_s$,
the only non-vanishing branches are the negative and positive ones, whose
coefficients reduce to
\begin{equation}
\begin{aligned}
A_r^{(-)}
&=
\binom{m}{r}\frac{\Delta^{\,n+m-r+1}}{n+m-r+1}
\\
&+
\delta_{r,n+m+1}\,
(-1)^n\frac{n!\,m!}{(n+m+1)!},    
\end{aligned}
\label{eq:Arminus_appendix_clean}
\end{equation}
and
\begin{equation}
\begin{aligned}
A_r^{(+)}
&=
\binom{n}{r}\frac{(-1)^r\Delta^{\,n+m-r+1}}{n+m-r+1}
\\
&+
\delta_{r,n+m+1}\,
(-1)^{n+1}\frac{n!\,m!}{(n+m+1)!}.    
\end{aligned}
\label{eq:Arplus_appendix_clean}
\end{equation}
We denote the length of the considered
span by \(L_s\), i.e., \(L_s=l_v\). Eq.~(\ref{eq:Cnm_final}) provides the closed-form representation
\begin{equation}
\begin{aligned}
I_{nm}^{(v,v)}
&=
\frac{1}{\beta_v^{n+m+2}}
\sum_{r=0}^{n+m+1}
[
A_r^{(-)}\bigl(M_r(0)-M_r(-\Delta_v)\bigr)
+
\\
&A_r^{(+)}\bigl(M_r(\Delta_v)-M_r(0)\bigr)
].
\end{aligned}
\label{eq:Cvv_Ar_Mr_app}
\end{equation}

Using the closed-form primitive derived in Appendix~\ref{app:Mr},
\begin{equation}
M_r(x)=\frac{1}{4\pi^2}\sum_{\ell=1}^4 \sigma_\ell\,H_r^{(\lambda_\ell)}(x),
\end{equation}
with
\begin{equation}
\lambda_1=4\pi^2ad,\quad
\lambda_2=4\pi^2ac,\quad
\lambda_3=4\pi^2bc,\quad
\lambda_4=4\pi^2bd,\nonumber
\end{equation}
and
\begin{equation}
(\sigma_1,\sigma_2,\sigma_3,\sigma_4)=(-1,+1,-1,+1),\nonumber
\end{equation}
Eq.~(\ref{eq:Cvv_Ar_Mr_app}) becomes
\begin{equation}
\begin{aligned}
I_{nm}^{(v,v)}
&=
\frac{1}{4\pi^2\beta_v^{n+m+2}}
\sum_{\ell=1}^4 \sigma_\ell
\sum_{r=0}^{n+m+1}
\\
&\Big[
A_r^{(-)}
\bigl(H_r^{(\lambda_\ell)}(0)-H_r^{(\lambda_\ell)}(-\Delta_v)\bigr)
\\
&
+
A_r^{(+)}
\bigl(H_r^{(\lambda_\ell)}(\Delta_v)-H_r^{(\lambda_\ell)}(0)\bigr)
\Big].
\end{aligned}
\label{eq:Ivv_after_Mr_sub_appendix}
\end{equation}

For $r=0$, one has
\begin{equation}
H_0^{(\lambda)}(x)=J(\lambda x),\nonumber
\end{equation}
hence
\begin{equation}
H_0^{(\lambda_\ell)}(\Delta_v)-H_0^{(\lambda_\ell)}(0)=J(\lambda_\ell\Delta_v),\nonumber
\end{equation}
and
\begin{equation}
H_0^{(\lambda_\ell)}(0)-H_0^{(\lambda_\ell)}(-\Delta_v)
=
-J(-\lambda_\ell\Delta_v)
=
J(\lambda_\ell\Delta_v),\nonumber
\end{equation}
where the odd parity of $J$ has been used.

For $r\ge 1$, ~\ref{app:Mr} gives
\begin{equation}
H_r^{(\lambda)}(x)
=
\frac{1}{\lambda^r}
\left[
\frac{(\lambda_\ell x)^r}{r}\,\mathrm{SI}(\lambda_\ell x)-\frac{1}{r}\,S_{r-1}(\lambda_\ell x)
\right].
\end{equation}
Define
\begin{equation}
\begin{aligned}
\mathcal{J}_r(L,\lambda)
&=
\delta_{r,0}\,L\lambda\,
{}_2F_3\!\left(
\frac{1}{2},\frac{1}{2};
\frac{3}{2},\frac{3}{2},\frac{3}{2};
-\frac{(L\lambda)^2}{4}
\right)
\\
&+
\bigl(1-\delta_{r,0}\bigr)
\left[
\frac{L^r}{r}\,\mathrm{SI}(L\lambda)-\frac{1}{r}\,S_{r-1}(L,\lambda)
\right].    
\end{aligned}
\label{eq:Jr_appendix_clean}
\end{equation}
We rewrite
\begin{equation}
\begin{aligned}
&\lambda_1'=\beta ad,\qquad
\lambda_2'=\beta ac,\qquad
\lambda_3'=\beta bc,\qquad
\lambda_4'=\beta bd,
\\
&\beta=4\pi^2\beta_v,    \nonumber
\end{aligned}
\end{equation}
and obtain for all $r\ge 0$
\begin{equation}
H_r^{(\lambda_\ell)}(\Delta_v)-H_r^{(\lambda_\ell)}(0)
=
\beta_v^r\,\mathcal{J}_r(L_s,\lambda_\ell'),
\end{equation}
and
\begin{equation}
H_r^{(\lambda_\ell)}(0)-H_r^{(\lambda_\ell)}(-\Delta_v)
=
(-1)^r\beta_v^r\,\mathcal{J}_r(L_s,\lambda_\ell'),
\end{equation}
where the parity properties of $\mathrm{SI}$ and $S_{r-1}$ have been used.

Substituting these endpoint differences into
Eq.~(\ref{eq:Ivv_after_Mr_sub_appendix}), we get
\begin{equation}
\begin{aligned}
I_{nm}^{(v,v)}
&=
\frac{1}{4\pi^2\beta_v^{n+m+2}}
\sum_{\ell=1}^4 \sigma_\ell
\\
&\sum_{r=0}^{n+m+1}
\beta_v^r
\Big[
(-1)^rA_r^{(-)}+A_r^{(+)}
\Big]
\mathcal{J}_r(L_s,\lambda_\ell').
\end{aligned}
\label{eq:Ivv_after_endpoint_appendix}
\end{equation}

Substituting Eqs.~\eqref{eq:Arminus_appendix_clean} and
\eqref{eq:Arplus_appendix_clean} into
Eq.~(\ref{eq:Ivv_after_endpoint_appendix}), the Kronecker-delta terms cancel
identically, and after using $\Delta=\beta_vL_s$ one obtains
\begin{equation}
\begin{aligned}
I_{nm}^{(v,v)}
&=
\frac{1}{4\pi^2\beta_v}
\sum_{\ell=1}^4 \sigma_\ell
\Bigg[
\sum_{i=0}^{m}
\binom{m}{i}
\frac{I_{i,m+n-i+1}(L_s,\lambda_\ell')}{m+n-i+1}
\\
&\qquad\qquad\qquad
+
\sum_{i=0}^{n}
\binom{n}{i}
\frac{I_{i,n+m-i+1}(L_s,\lambda_\ell')}{n+m-i+1}
\Bigg],
\end{aligned}
\label{eq:Ivv_preT_appendix_clean}
\end{equation}
where
\begin{equation}
I_{p,q}(L,\lambda)
=
\begin{cases}
\displaystyle
&\mathrm{SI}(L\lambda)\,L^{p+q}\frac{(p-1)!\,q!}{(p+q)!}
\\
&-
\sum_{r=0}^{q}
\binom{q}{r}
\frac{(-1)^rL^{q-r}}{p+r}S_{p+r-1}(L,\lambda),
 p\ge 1,
\\[1.0em]
\displaystyle
&\sum_{r=0}^{q}
\binom{q}{r}(-1)^rL^{q-r}\,\mathcal{J}_r(L,\lambda),
 p=0.
\end{cases}
\label{eq:Ipq_appendix_clean}
\end{equation}

We then define
\begin{equation}
T_{a,b}(L,\lambda)
=
\sum_{i=0}^{a}
\binom{a}{i}
\frac{I_{i,a+b-i+1}(L,\lambda)}{a+b-i+1}.
\label{eq:Tab_appendix_clean}
\end{equation}
Therefore, the kernel $I_{nm}^{(v,v)}$
obtained in Eq.~(\ref{eq:Ivv_preT_appendix_clean}) directly reduces to the conventional single-span closed-form kernel $\kappa_x$:
\begin{equation}
\begin{aligned}
\kappa_{x} &=
\displaystyle \frac{1}{\beta}\sum_{k=1}^{4}(-1)^k \Big[T_{m,n}\!\big(L_s,\lambda_k'\big)
+ T_{n,m}\!\big(L_s,\lambda_k'\big) \Big]\\
T_{a,b} \big(L, \lambda\big) & = \sum_{i=0}^{a} \binom{a}{i}\frac{I_{i,a+b-i+1}\big(L,\lambda\big)}{a+b-i+1}, \qquad \beta = 4\pi^2 \beta_{2,\mathrm{eff}}\\[0.1em]
I_{p,q}(L,\lambda) & = \begin{cases} \displaystyle \mathrm{SI}(L\lambda) \cdot L^{p+q} \cdot \frac{(p-1)!\,q!}{(p+q)!}\\
\displaystyle - \sum_{r=0}^{q} \binom{q}{r} \frac{(-1)^r L^{q-r}}{p+r} S_{p+r-1}(L,\lambda), & p\geq 1 \\[0.1em]
\displaystyle \sum_{r=0}^{q} \binom{q}{r} (-1)^r L^{q-r}\, \mathcal{J}_r(L,\lambda), & p = 0 \end{cases} \\[0.em]
 \mathcal{J}_r(L,\lambda) &=\delta_{r,0} \cdot L\lambda \cdot {}_2F_3\!\left(\frac{1}{2},\frac{1}{2};\frac{3}{2},\frac{3}{2},\frac{3}{2};-\frac{(L\lambda)^2}{4}\right) \\
&\qquad + \big(1-\delta_{r,0} \big) \left[\frac{L^r}{r}\,\mathrm{SI}(L\lambda)-\frac{1}{r}\,S_{r-1}(L,\lambda)\right] \\[0.em]
 S_{k}(L,\lambda) &= \sum_{p=0}^{k} \frac{(-1)^p\,k!\,L^{k-p}}{(k-p)!\,\lambda^{p+1}} \sin\!\left(L\lambda-\frac{\pi}{2}(p+1)\right) \\
&\qquad + \frac{k!}{\lambda^{k+1}} \sin\!\left(\frac{\pi}{2}(k+1)\right) \\[0.em]
\lambda_1' &=\beta a d, \quad \lambda_2'=\beta a c, \quad \lambda_3'=\beta b c, \quad \lambda_4'=\beta b d.
\end{aligned}
\label{eq:kappa_final_appendix_clean}
\end{equation}

In the special case where the effective dispersion vanishes, i.e., $\beta_{2,\mathrm{eff}}=0$, the kernel degenerates and the frequency integral reduces to a constant. As a result, the core integral admits a much simpler closed-form expression, which can be directly obtained as:

\begin{equation}
\kappa_x = (b-a)(d-c)\frac{L^{n+m+2}}{(n+1)(m+1)}, \qquad \beta_{2,\mathrm{eff}}=0.
\end{equation}

This expression corresponds to the limit of the general formulation when $\beta_{2,\mathrm{eff}}$→0.

\end{appendices}

\end{document}